\documentclass[%
 aip,
 amsmath,amssymb,
 reprint,%
]{revtex4-1}

\usepackage{romannum}
\usepackage{graphicx}%
\usepackage{dcolumn}%
\usepackage{bm}%
\usepackage[utf8]{inputenc}
\usepackage[T1]{fontenc}
\usepackage{enumitem}
\usepackage{graphicx}

\usepackage{pgf}
\usepackage{tikz}
\usetikzlibrary{arrows,automata}
\usepackage{verbatim}

\usepackage[textwidth=4cm]{todonotes}

\newcommand{\abs}[1]{\vert#1\vert}

\newcommand{\FF}{\mathcal{F}}
\newcommand{\II}{\mathcal{I}}
\newcommand{\DD}{\mathcal{D}}

\usepackage{amsthm}
\newtheorem{lemma}{Lemma}

\newtheorem{prop}{Proposition}

\usepackage{tikz}
\newcommand\copyrighttext{%
  \footnotesize This article may be downloaded for personal use only. Any other use requires prior permission of the author and AIP Publishing. The following article appeared in Chaos \textbf{29}, 123119 (2019) and may be found at https://doi.org/10.1063/1.5122739 .}
\newcommand\copyrightnotice{%
\begin{tikzpicture}[remember picture,overlay]
\node[anchor=south] at (current page.south) {\fbox{\parbox{\dimexpr\textwidth-\fboxsep-\fboxrule\relax}{\copyrighttext}}};
\end{tikzpicture}%
}

\begin{document}

\pagenumbering{arabic}

\title{Multistability in lossy power grids and oscillator networks}

\author{Chiara Balestra}%
 \email{chiara.balestra@edu.unito.it}
  \affiliation{Institute for Theoretical Physics, University of Cologne, K\"oln, 50937, Germany}
  \affiliation{Department of Mathematics "Giuseppe Peano", 
  Università degli Studi di Torino, 10123 Torino, Italy}

\author{Franz Kaiser}%
 \email{f.kaiser@fz-juelich.de}
  \affiliation{Institute for Theoretical Physics, University of Cologne, K\"oln, 50937, Germany}
  \affiliation{Forschungszentrum J\"ulich, Institute for Energy and Climate Research (IEK-STE), 52428 J\"ulich, Germany}

\author{Debsankha Manik}%
 \email{debsankha.manik@ds.mpg.de}
  \affiliation{Max Planck Institute for Dynamics and Self-Organization, Am Faßberg 17, 37077 Göttingen, Germany}

\author{Dirk Witthaut}%
 \email{d.witthaut@fz-juelich.de}
  \affiliation{Institute for Theoretical Physics, University of Cologne, K\"oln, 50937, Germany}
  \affiliation{Forschungszentrum J\"ulich, Institute for Energy and Climate Research (IEK-STE), 52428 J\"ulich, Germany}

\date{\today}

\begin{abstract}
Networks of phase oscillators are studied in various contexts, in particular in the modeling of the electric power grid. A functional grid corresponds to a stable steady state, such that any bifurcation can have catastrophic consequences up to a blackout. But also the existence of multiple steady states is undesirable, as it can lead to  transitions or circulatory flows. Despite the high practical importance there is still no general theory of the existence and uniqueness of steady states in such systems. Analytic results are mostly limited to grids without Ohmic losses. In this article, we introduce a method to systematically construct the solutions of the real power load-flow equations in the presence of Ohmic losses and explicitly compute them for tree and ring networks. We investigate different mechanisms leading to multistability and discuss the impact of Ohmic losses on the existence of solutions.
\end{abstract}

\maketitle
\copyrightnotice

\begin{quotation}
The stable operation of the electric power grid relies on a precisely synchronized state of all generators and machines. All machines rotate at exactly the same frequency with fixed phase differences leading to steady power flows throughout the grid. Whether such a steady state exists for a given network is of eminent practical importance. The loss of a steady state typically leads to power outages up to a complete blackout. But also the existence of multiple steady states is undesirable, as it can lead to sudden transitions, circulating flows and eventually also to power outages. Steady states are typically calculated numerically, but this approach gives only limited insight into the existence and (non-)uniqueness of steady states. Analytic results are available only for special network configurations, in particular for grids with negligible Ohmic losses or radial networks without any loops. In this article, we introduce a method to systematically construct the solutions of the real power load-flow equations in the presence of Ohmic losses. We calculate the steady states explicitly for elementary networks demonstrating different mechanisms leading to multistability. Our results also apply to models of coupled oscillators which are widely used in theoretical physics and mathematical biology.  
\end{quotation}

\section{Introduction}

The electric power grid is one of the largest man-made systems, and a stably operating grid is integral for the entire economy, industry, and almost all other technical infrastructures. The complexity of the power grid with thousands of generators, substations and transmission elements calls for an interdisciplinary approach to ensure stability in a transforming energy system  \cite{brummit2013,timme2015}. In particular, the interrelation of structure and stability of complex grids has received widespread attention in recent years, see e.g.~\cite{filatrella2008analysis,witthaut2012,dorfler2012synchronization,witthaut2013,dorfler2013synchronization,manik2014,simpson2017a,simpson2017b,jafarpour2019multistable}. These endeavours have been aided by the similarity of mathematical models across scientific disciplines. The fundamental models for power grid dynamics such as the classical model or the structure-preserving model \cite{anderson1977power,bergen1981structure} are mathematically equivalent to the celebrated Kuramoto model with inertia   \cite{kuramoto1975international,strogatz2000kuramoto,acebron2005kuramoto,arenas2008synchronization}. Therefore, results obtained on networks of Kuramoto oscillators can be easily translated to power grids and vice versa.

A central question across disciplines is whether a stable steady state exists and whether it is unique given a certain network structure. In the context of power grids, it is desirable to have a unique steady state. Grid operators strive to maintain the flows across each line below a certain limit to avoid disruptions. Ensuring this is much more difficult if one has to take into account multiple steady states, and hence multiple flow patterns across the lines. Analytic results have been obtained for various special cases. In particular, multistability has been ruled out for lossless grids in the two limiting cases of very densely connected networks \cite{kuramoto1975international,taylor2012there} as well as tree-like networks (very sparse) \cite{manik2017}. The existence of a steady state is determined by two factors: the distribution of the real power injections (natural frequencies for Kuramoto oscillators) and the strength of connecting lines. A variety of related results have been obtained for tree-like distribution grids in power engineering, see e.g.~\cite{chiang1990}. 

The situation is more involved for networks of intermediate sparsity such as power transmission grids, which can give rise to multistability \cite{korsak1972question,wiley2006size,ochab2010synchronization,delabays2016multistability,manik2017,delabays2017,jafarpour2019multistable}. The existence of multiple steady states in meshed networks can be traced back to the existence of \emph{cycle flows} that do not affect the power balance at any node in the grid. The number of and size of the cycles in the grid is thus an essential factor that determines the number of steady states \cite{manik2017}. Exploring the quantitative relationship between these topological factors and multistability, rigorous bounds on the number of steady states and mechanisms for a grid to switch from one steady state to another one have been found \cite{ochab2010synchronization,mehta2015algebraic,delabays2016multistability,coletta2016topologically,manik2017,delabays2017,jafarpour2019multistable}.

Despite the great theoretical progress, a general theory of the solvability of the power flow equations is still lacking. Most analytic studies focus on lossless grids \cite{korsak1972question,wiley2006size,ochab2010synchronization,dorfler2013synchronization,delabays2016multistability,manik2017,delabays2017,simpson2017a,jafarpour2019multistable,mehta2015algebraic,park2019monotonicity,zachariah2018distributions} or tree-like grids \cite{simpson2017b,chiang1990,miu2000,lavaei2012,bolognani2015}. Analytic results are extremely rare for the full power flow equations with Ohmic losses in meshed networks \cite{korsak1972question, bukhsh2013local,cui2019solvability}.  

In this article, we present a new approach to compute the steady states of the real power flow equations in general networks in the presence of Ohmic losses, extending a prior study of lossless grids \cite{manik2017}. Our main contribution is a stepwise procedure to construct solutions. In a first step, flows and losses are treated as independent variables, turning the load flow equations into a linear set of equations. The inherent relationship between flows and losses is reintroduced in a second step. Choosing an appropriate basis for the solution space of the linear set of equations, we can explicitly compute the coefficients that lead to a consistent solution. Using this approach, we show that Ohmic losses in general have two contrary effects on the solvability of the real power flow equations: On the one hand, increasing losses requires higher line capacities to be able to transport the same amount of power thereby potentially destabilizing the grid and thus losing stable fixed points. On the other hand, we demonstrate for two very basic topologies that high line losses may also cause multistability leading to additional stable fixed points through a mechanism non-existent for the lossless case. 

The article is organized as follows: we first specify the mathematical structure of the problem and fix the notation in section \ref{sec:grid}. We then briefly review the lossless case in section \ref{sec:lossless} to illustrate the fundamental importance of cycles and cycle flows. Section \ref{sec:withloss} constitutes the main part of the paper, introducing the stepwise approach. We then investigate two topologies in detail: a tree and a ring network, for which we lay down the procedures for computing all the steady states, in sections \ref{sec:tree} and \ref{sec:cycle}. 

\section{Steady states in power grids and oscillator networks}
\label{sec:grid}

The load-flow equations constitute the fundamental model to describe the steady state of an AC power grid. The system state is defined in terms of the magnitude and phase of the nodal voltages $V_j e^{i \theta_j}$, $j\in\{1,\ldots,N\}$, that have to satisfy the energy conservation law. The nodes provide or consume a certain amount of real power $P_j$ such that the real power balance reads
\begin{align}
   P_j = \sum_k & \Big(
              b_{jk} V_j V_k \sin(\theta_j - \theta_k) \nonumber \\
             & \quad + g_{jk} \left(V_j^2 - V_j V_k \cos(\theta_j - \theta_k) \right)\Big). 
   \label{eq:realpower-intro}
\end{align}
Here, $g_{jk}$ is the conductance of the line $(j,k)$, while the susceptance is given by $-b_{jk}$ (not $+b_{jk}$!). By this definition both $g_{jk}$ and $b_{jk}$ are generally positive for all transmission elements, with $g_{jk}=b_{jk}=0$ if the two nodes $j$ and $k$ are not connected. The variation of the voltage magnitudes $V_j$ is intimately related with the provision and demand for reactive power. In general, generator nodes adapt the reactive power to fix the voltage to the reference level $V_j = V_{\rm ref}$, while load nodes consume a fixed value of reactive power. The voltage magnitude $V_j$ can depart from the reference level \cite{simpson2016}, but strict security rules are imposed to limit this voltage variation. In the present article we will focus on the real power balance equation (\ref{eq:realpower-intro}) to explore the existence of solutions and possible routes to multistability. We neglect voltage variability to reduce the complexity of the problem and refer to \cite{simpson2017a,simpson2017b} for a detailed discussion of this issue. Technically, this corresponds to the assumption that the reactive power can be balanced at all nodes. Using appropriate units, referred to as the pu system in power engineering \cite{wood2013} we can thus set 
\begin{equation*}
   V_j = V_{\rm ref} = 1 \\
\end{equation*} 
for all nodes.
 
The network structure plays a decisive role for the existence and stability of steady states. This structure is encoded in the coupling coefficients $b$ and $g$. For a given transmission line $(j,k)$ with resistance $r_{jk}$ and reactance $x_{jk}$ we have
\begin{equation}
    g_{jk} - i b_{jk} = \frac{1}{r_{jk} + i x_{jk}}.
\end{equation} 
 In high voltage transmission grids, Ohmic losses are typically small such that $g$ is small compared to $b$. In the limit of a lossless line, we obtain $g_{jk} = 0$ and $b_{jk} = 1/x_{jk} > 0$. In contrast, $b$ and $g$ are of similar magnitude in distribution grids.

A mathematically equivalent problem arises in the analysis of steady states of dynamical power system models. In particular, the dynamics of coupled synchronous machines is determined by the swing equation \cite{nishikawa2015}
\begin{equation}
   M_j \frac{d^2 \theta_j}{dt^2} + D_j \frac{d \theta_j}{dt} = P_j - P_j^{\rm el},
   \label{eq:swing}
\end{equation} 
whose steady states are again determined by Eq.~\eqref{eq:realpower-intro}. Furthermore, coupled oscillator models are used to describe the collective motion of various systems across scientific disciplines. For instance, the celebrated Kuramoto model considers a set of $N$ limit cycle oscillators whose state is described by their phases $\theta_j$ along the cycle. In many important applications\cite{daido1992,abrams2008,14hamilton}, the equations of motions of the coupled system are given by 
\begin{equation}
   \frac{d \theta_j}{dt} = \omega_j + \sum_{k=1}^N K_{jk} 
      \sin(\theta_k - \theta_j + \gamma_{jk}),
      \label{eq:Kuraphaselags}
\end{equation} 
where $\omega_j$ is the intrinsic frequency of the $j$-th oscillator, $K_{jk} = K_{kj}$ is the coupling strength of the link between oscillators $j$ and $k$ and $\gamma_{jk} =\gamma_{kj}$ is a phase shift. The fixed points of this model are determined by the algebraic equations $d\theta_j/dt = 0$ that are cast into the following form by using basic trigonometric identities
\begin{align}      
  \omega_j + \sum_k & K_{jk} \sin(\gamma_{jk}) = 
  \sum_k \Big( K_{jk} \cos(\gamma_{jk}) \sin(\theta_j^* - \theta_k^*) \nonumber \\
    & + K_{jk} \sin(\gamma_{jk}) \left[ 1 - 
    \cos(\theta_j^* - \theta_k^*) \right]\Big),
    \label{eq:Kuramoto_fixed}
\end{align} 
 where $\vec{\theta}^*=(\theta_1^*,\ldots,\theta_N^*)$ is a fixed point. This equation is identical to the real power balance (\ref{eq:realpower-intro}) if we identify
$P^{\rm in}_j = \omega_j + \sum_k K_{jk} \sin(\gamma_{jk})$, $b_{jk} = K_{jk} \cos(\gamma_{jk})$ and $g_{jk} = K_{jk} \sin(\gamma_{jk})$. We note that in the limit of a lossless line, $\gamma_{jk}=0$ for all edges. In the following, we will fix a slack node $s$ that can provide an infinite amount of power $P_s$, which translates as an additional free parameter to the Kuramoto model given by the frequency at the node corresponding to the slack node $\omega_s$. Therefore, different fixed points, i.e., solutions to Eq.~\eqref{eq:Kuramoto_fixed}, can have a different frequency at the slack node $\omega_s$ in this set-up, which differs from the way fixed points are typically considered in the Kuramoto model.

The stability of a given fixed point $\vec\theta^*$ is assessed by adding a small perturbation\cite{strogatz2018} and then using linear stability analysis,
\begin{equation}
   \theta_j = \theta_j^* + \xi_j, \qquad j=1,\ldots,N.
\end{equation}
For the first order model, the dynamics of the perturbation is to linear order given by
\begin{align*}
    \frac{d \xi_j}{dt} &= \sum_{k=1}^N w_{jk} (\xi_k - \xi_j)
\end{align*}
with the weights
\begin{align*}
   w_{jk} &= K_{jk} \cos(\theta_k^*-\theta_j^* + \gamma_{jk}) \nonumber \\
          &= b_{jk} \cos(\theta_k^*-\theta_j^*) -
             g_{jk} \sin(\theta_k^*-\theta_j^*).
\end{align*}

This relation is expressed in vectorial form as
\begin{equation}
   \frac{d \vec \xi}{dt} = - \boldsymbol\Lambda \vec \xi
\end{equation}
with the Laplacian matrix $\boldsymbol\Lambda \in \mathbb{R}^{N \times N}$ with elements
\begin{equation}
   \Lambda_{jk} = \left\{ \begin{array}{l l l}
   - w_{jk} &   \mbox{  for} \; & j \neq k \\
   \sum \nolimits_\ell w_{j\ell} &   \mbox{  for}\;&j = k.
   \end{array} \right.
   \label{eq:def-Laplace}
\end{equation}
Before we proceed we note that $\boldsymbol\Lambda$ always has a zero eigenvalue corresponding to a global shift of all phases $\theta_j \rightarrow \theta_j + c$ that does not affect the synchronization of the system. We thus discard this mode and limit the stability analysis to the subspace perpendicular to it
\begin{equation}
   \DD_\perp = \{ \vec y \in \mathbb{R}^{N} | (1,1,\ldots,1) \vec y^\top = 0 \}.
\end{equation}
A steady state is linearly stable if all perturbations in $\DD_\perp $ are damped exponentially, which is the case if the real part of all eigenvalues of $\boldsymbol\Lambda$ are strictly positive (except for the zero eigenvalue corresponding to a global phase shift).

Stability analysis becomes rather simple in the lossless case. Assuming that the network is connected and that the phase differences along any line are limited as
\begin{equation}
    |\theta_k^*-\theta_j^*| < \frac{\pi}{2},
    \label{eq:normalop}
\end{equation}
the matrix $\boldsymbol\Lambda$ is a proper graph Laplacian of an undirected graph, whose relevant eigenvalues are always positive. Hence, Eq.~(\ref{eq:normalop}) is a sufficient condition for linear stability but not a necessary one. Stable steady states that violate condition (\ref{eq:normalop}) do exist at the boundary of the stability region, but in most cases states with phase differences that are this large are unstable \cite{manik2014,chen2016,delabays2017}. Hence, we typically focus on states that do satisfy (\ref{eq:normalop}) and refer to this as the \emph{normal operation} of the grid \cite{manik2017}.

The stability analysis is more involved in the presence of Ohmic losses, as $\boldsymbol\Lambda$ is no longer symmetric. Hence, it rather corresponds to the Laplacian of a directed network, whose stability is harder to grasp analytically. In this case we will evaluate the linear stability of different steady states by direct numerical computations.

However, in the case where all off-diagonal elements of this matrix are strictly negative, we are able to gain limited analytical insight by the following Lemma:
\begin{lemma}
\label{lem:stability_condition}
Let $\vec{\theta}^*\in\mathbb{R}^N$ be an equilibrium of the Kuramoto model with phase lags as defined in Eq.~\eqref{eq:Kuraphaselags}. The equilibrium is linearly stable if all edges $(j,k)$ have positive weights
\begin{align*}
    w_{jk}=K_{jk} \cos(\theta_k^*-\theta_j^* + \gamma_{jk})>0,\quad\forall (j,k).
\end{align*}
\end{lemma}

A proof is given in Appendix~\ref{sec:proof_lemma_1}. Note that similar results have also been reported in Ref.~\cite{schiffer_synchronization_2013}.

\section{The lossless case}
\label{sec:lossless}

We briefly review the analysis of the lossless case to introduce the fundamentals of our approach as well as some notation and methodology following Ref.~\cite{manik2017}.

\subsection{Constructing solutions}

Consider a graph $G$ consisting of $N$ nodes and $M$ edges. The lossless case is recovered from equation (\ref{eq:realpower-intro}) by putting $g_{jk}=0$ and assuming $V_j = V_{\rm ref} = 1, \forall j$. The steady states are then determined by the equation
\begin{equation}  
   \vec P=  \boldsymbol I \boldsymbol B_d \sin(\boldsymbol I^\top\vec{\theta}) \, .
     \label{eq:p-lossless}
\end{equation} 
Here, the sine function is assumed to be taken element-wise and we summarized all quantities in a vectorial form
\begin{align*}
        \vec P &= (P_1, \ldots, P_N )^\top
        \in \mathbb{R}^{N} \, ,  \\
    \vec \theta &= (\theta_1, \ldots, \theta_N )^\top
        \in \mathbb{R}^{N} \, ,  \\
    \boldsymbol B_d &= \mathrm{diag}(b_1,\ldots,b_M)\in\mathbb{R}^{M\times M}\, .
\end{align*}
The topology of the network is encoded in the node-edge incidence matrix $\boldsymbol I \in \mathbb{R}^{N \times M}$ with elements \cite{newman2010}
\begin{equation}
   I_{j,e} = \left\{
   \begin{array}{r l }
      +1 & \; \mbox{if node $j$ is the tail of edge $e  \,  \hat{=} \, (j,\ell)$},  \\
      -1 & \; \mbox{if node $j$ is the head of edge $e  \,  \hat{=} \, (j,\ell)$},  \\
      0     & \; \mbox{otherwise}.
  \end{array} \right.
\end{equation}
Based on this matrix, we also fix an orientation for each of the network's edges \cite{godsil2013algebraic}. Steady states exists only if the power injections of the entire grid are balanced, i.e., $\sum_j P_j = 0$, which we assume to hold. 

The main idea to construct all solutions of Eq.~\eqref{eq:p-lossless} is to shift the focus from nodal quantities to edges and cycles of the network. To do so, we define a vector $\vec F = (F_1,\ldots, F_M)^\top \in \mathbb{R}^M$ of flows on the network's edges
\begin{equation}  
   \vec F = \boldsymbol B_d  \sin(\boldsymbol I^\top \vec \theta).
   \label{eq:FfromT-lossless}
\end{equation} 
If a component of the flow vector is larger than zero, $F_e > 0$, the flow on link $e=(k,j)$ is directed from $k$ to $j$ and if $F_e < 0$ from $j$ to $k$. Therefore $F_e$ physically denotes the flow from the tail of the edge $e$ to the head of $e$. Eq.~\eqref{eq:p-lossless} then becomes
\begin{equation}  
   \vec P=  \boldsymbol I \vec F \, .
     \label{eq:lin-lossless}
\end{equation} 

Solutions of Eq.~\eqref{eq:p-lossless} may be constructed by first solving Eq.~\eqref{eq:lin-lossless} and then rejecting all solution candidates which are incompatible with Eq.~\eqref{eq:FfromT-lossless}. Solutions of (\ref{eq:lin-lossless}) may be obtained based on the following observation: the kernel of the incidence matrix $\boldsymbol I$ corresponds exactly to \emph{cycle flows}, a cycle flow referring to a constant flow along a cycle with no in- or out-flow\cite{ronellenfitsch2016,ronellenfitsch2017,horsch2018}. The kernel has dimension $M-N+1$, which reflects the fact that the cycles in a graph forms a vector space of dimension $M-N+1$ \cite{diestel2010}, a basis set of this space is called a fundamental cycle basis. A set of fundamental cycles $B$ is encoded in the corresponding cycle-edge incidence matrix $\boldsymbol C^B \in \mathbb{R}^{M \times (M-N+1)}$ with elements
\begin{equation}
   C^B_{e,c} = \left\{
   \begin{array}{r l }
      +1 & \; \mbox{if the edge $e$ is part of the cycle $c$}  \\
      -1 & \; \mbox{if the reversed edge $e$ is part of cycle  $c$}  \\
      0     & \; \mbox{otherwise}.
  \end{array} \right.
\end{equation}

Then, all solutions of equation (\ref{eq:lin-lossless}) can be written as
\begin{equation} 
   \vec F = \vec F^{(s)} + \boldsymbol C^B \vec f,
\end{equation} 
where $\vec F^{(s)} \in \mathbb{R}^M$ is a specific solution and $\vec f \in \mathbb{R}^{M-N+1}$ gives the strength of the cycle flows along each cycle in the chosen cycle basis. Having obtained a flow vector $\vec F$, we can simply construct the associated phases as follows. Start at the slack node $s$ and set $\theta_s = 0$. Then proceed to a neighbouring node $j$. Assuming that the connecting edge $e\hat{=}(j,s)$ is oriented from node $s$ to node $j$, the phase value reads
\begin{equation}
    \theta_j = \theta_s + \Delta_{e},
    \label{eq:theta_incremental}
\end{equation}
where the phase difference $\Delta_{e}$ is reconstructed from the flow $F_{e} $ by inverting Eq.~\eqref{eq:FfromT-lossless},
\begin{align}
     \Delta_{e}^+ = \arcsin(F_{e}/b_{e})  
    \quad \mbox{or} \nonumber\\  
    \Delta_{e}^- = \pi - \arcsin(F_{e}/b_{e}). 
    \label{eq:Delta_pm}
\end{align}

For each edge $e$ we have to decide whether we take the $+$-solution or the $-$-solution in Eq.~\eqref{eq:Delta_pm}. To keep track of this choice, we decompose the edge set  of the network $E$ into two parts,
\begin{align*}
   E_+ &= \{ e \in E \mid \Delta_e =  \Delta_e^+ \} \\
   E_-  &= \{ e \in E \mid \Delta_e =  \Delta_e^- \},
\end{align*}
such that $E = E_+ \cup E_-$. Not all solutions obtained this way are physically correct. We can obtain the physically correct ones by making sure that the sum of the phase differences around any fundamental cycle yields zero or an integer multiple of $2 \pi$, which is equivalent to the winding numbers
\begin{equation}
    \varpi_c = \frac{1}{2\pi} \sum_{e=1}^M C^B_{e,c} \Delta_e^{\pm},
    \label{eq:geom_condition}
\end{equation}
summarized in the vector $\vec \varpi = (\varpi_1,\ldots,\varpi_{M-N+1})^\top$ being integer $\varpi_c \in \mathbb{Z}$. It should be noted that the choice $\Delta_e^+$ corresponds to the state of normal operation discussed in section \ref{sec:grid}. Hence, states with $E_- = \emptyset$ are guaranteed to be stable, while states with $E_- \neq \emptyset$ are typically (but not always) unstable \cite{manik2014,delabays2017,manik2017}. We summarize these results in the following proposition due to Ref.~\cite{manik2017}. 

\begin{prop}
\label{th:fp-dyn-geo}
Consider a connected lossless network with power injections $\vec P \in \mathbb{R}^{N}$. Then the following two statements are equivalent:
\begin{enumerate}
\item $\vec \theta$ is a steady state, i.e., a real solution of equation (\ref{eq:p-lossless}). 
\item The flows $\vec F \in \mathbb{R}^M$ satisfy the `dynamic' conditions (\ref{eq:lin-lossless}) with $|F_e| \le b_e$ such that
\begin{equation}
    \vec F = \vec F^{(s)} + \boldsymbol C^B \vec f
\end{equation}
and the geometric condition \eqref{eq:geom_condition}
\begin{equation}   
    \vec \varpi(\vec f) \in \mathbb{Z}^{M-N+1}.
\end{equation}
for some decomposition $E = E_+ \cup E_-$.
\end{enumerate}
\end{prop}

\section{Power Grids with Ohmic losses}
\label{sec:withloss}

We now extend the approach introduced above to power grids with Ohmic losses or oscillator networks with a general trigonometric coupling. The steady states are determined by the real power balance equation  (cf.~Eq.~\eqref{eq:realpower-intro})
\begin{equation}
   P_j = \sum_{k=1}^N\Big( b_{jk} \sin(\theta_j - \theta_k)  + g_{jk} \left[1 -  \cos(\theta_j - \theta_k)\right]\Big). 
   \label{eq:balance}
\end{equation}
Before we proceed to construct the solution to these equations we note an important difference to the lossless case. The Ohmic losses occurring on the lines are not a priori known as they depend on the phases $\theta_1,\ldots,\theta_N$. Hence the real power balance for the entire grid now reads
\begin{equation}
   \sum_{j=1}^N P_j = P_{\rm losses}(\theta_1,\ldots,\theta_N).
\end{equation}
Thus for arbitrary $P_1,\ldots,P_N $, there will typically be no solution. This issue is solved by assuming that one of the nodes, referred to as the slack node, can provide an arbitrary amount of power to balance the losses. For the sake of consistency, we label the slack as $j=1$ throughout this article and set $\theta_1=0$. We note that the choice of a particular slack node is often arbitrary. In transmission grids one typically chooses a node with high generation, whereas in distribution grids one can choose the connection to the higher grid level. Other approaches using a distributed slack bus also exist, see e.g.~\cite{tong_wu_locational_2005}. 

To solve the set of equations (\ref{eq:balance}) for the remaining nodes $j\in\{2,\ldots,N\}$ we decompose it into different parts as before and first formulate a linear system  of equations. Before we start, we fix some notations by defining the \emph{unsigned} incidence matrix $\boldsymbol E \in \mathbb{R}^{N \times M}$ with elements $E_{j,e} = |I_{j,e}|$. For each edge $e \hat{=} (j,k)$ we now define the losses by
\begin{equation*}
   L_e = g_{e} \left[1 -  \cos(\theta_j - \theta_k)\right].
\end{equation*}

Using this notation, the power balance equations can be decomposed into three parts. First we have the dynamic condition, which now reads
\begin{equation}
    (\text{\Romannum{1}a})\;  P_j = \sum_{e=1}^M I_{j,e} F_e + E_{j,e} L_e 
    ,\quad \forall \, j\in\{2,\ldots,N\}.
    \label{eq:dyncon}
\end{equation}
Flows and losses are limited by the line parameters which is represented by the following conditions
\begin{equation}
  (\text{\Romannum{1}b})~ \frac{F_e}{b_e} \in [-1,1], \quad \frac{L_e}{g_e} \in [0,2]
         ,\quad \forall \, e\in\{1,\ldots,M\}.   
         \label{eq:linelimits} 
\end{equation}
In addition to that, flows and losses are not independent, but are both functions of the phase difference $\theta_j - \theta_k$. Using the trigonometric identity $\sin^2 + \cos^2 =1$ we obtain the flow-loss condition
\begin{equation}
   (\text{\Romannum{2}})~\left( \frac{F_e}{b_e} \right)^2 
       + \left( \frac{L_e}{g_e} - 1 \right)^2 = 1,
     \quad\forall \, e\in\{1,\ldots,M\}.
     \label{eq:flowloss}
\end{equation}
Finally, we have a geometric condition as in the lossless case
\begin{equation}
   (\text{\Romannum{3}})~\varpi_c(F_1,\ldots, L_1,\ldots) = z \quad 
   \mbox{with} \; z \in \mathbb{Z},
   ~ \forall \, \mbox{cycles } \, c.
   \label{eq:geo}
\end{equation}
In comparison to the lossless case we have $M$ additional degrees of freedom $L_1,\ldots, L_M $ and $M$ additional nonlinear conditions (\ref{eq:flowloss}) to fix them. Furthermore, the knowledge of both $F_e$ and $L_e$ is sufficient to fix the phases completely. Eq.~\eqref{eq:Delta_pm} is replaced by
\begin{align}
     \Delta_{e} = \left\{ \begin{array}{l l l}
     \arcsin(F_{e}/b_{e})  &\quad \mbox{ if } &
     L_e \le g_e \\
    \pi - \arcsin(F_{e}/b_{e}) & \quad \mbox{ if }&
    L_e > g_e.
    \end{array} \right.
    \label{eq:Delta-FL-loss}
\end{align}
Still, there are two solution branches $\pm$ per edge as in the lossless case, because the quadratic equation (\ref{eq:flowloss}) has two solutions in general. We summarize these findings in the following proposition.
\begin{prop}
\label{th:fp-dyn-lossy-case}
Consider a connected lossy network with power injections $\vec P \in \mathbb{R}^{N}$. Then the following two statements are equivalent:
\begin{enumerate}
\item $\vec \theta$ is a steady state, i.e., a real solution of equation (\ref{eq:balance}). 
\item The flows $\vec F \in \mathbb{R}^M$ and losses $\vec L \in \mathbb{R}^M$ 
satisfy the `dynamic' conditions (\ref{eq:dyncon}) with $|F_e| \le b_e$ and $0 \le L_e \le 2 g_e$, the flow loss condition (\ref{eq:flowloss}), and the geometric condition 
\begin{equation}   
    \vec \varpi \in \mathbb{Z}^{M-N+1}.
\end{equation}
\end{enumerate}
\end{prop}

To find actual solutions, we thus have to solve the linear set of equations (\ref{eq:dyncon}) subject to a variety of nonlinear constraints (\ref{eq:linelimits}-\ref{eq:geo}). Remarkably, this can be accomplished in an iterative fashion such that we find the following general strategy to construct solutions:
\begin{enumerate}
\item Construct the solution space of the linear set of equations (\ref{eq:dyncon}), which yields a potentially large set of solution \emph{candidates}. This set is gradually reduced in the further steps until only the actual solutions are left. 
\item Use the flow-loss condition (\ref{eq:flowloss}) to reduce the degrees of freedom of the system. In particular, all remaining solution candidates can be expressed in terms of the cycle flow strengths and a set of indices $\pm$ which indicate the solution branch for each edge.
\item Finally, fix the cycle flows by the geometric conditions (\ref{eq:geo}).
\end{enumerate}
Remarkably, we will see in the following that condition~(\ref{eq:linelimits}) on the line limits is automatically satisfied, if a real solution of the flow-loss condition (\ref{eq:flowloss}) exists, so we do not have to explicitly consider this condition (see Lemma~\ref{lem:rlalpha}). We further note that the addition of cycle flows still does not affect the power balance, so the cycle flows remain a basic degrees of freedom in equation (\ref{eq:dyncon}). The losses $L_e$ are fixed only in the second step using the flow-loss condition (\ref{eq:flowloss}). Hence, the resulting losses depend on the strength of cycle flows. We now illustrate this approach by explicitly constructing the solutions for a tree network and a single cycle. We will show that including losses gives rise to an additional mechanism of multistability. 

\section{Tree Networks}
\label{sec:tree}

We will first consider tree networks, i.e.~networks without any closed cycles. Hence, we do not have to take into account the geometric condition (\ref{eq:geo}) and focus on the solution of the flow-loss condition (\ref{eq:flowloss}). 

\subsection{Fundamentals}
\label{sec:lossy-tree}

\begin{figure}[t]
    \begin{center}
        \includegraphics[width=\columnwidth]{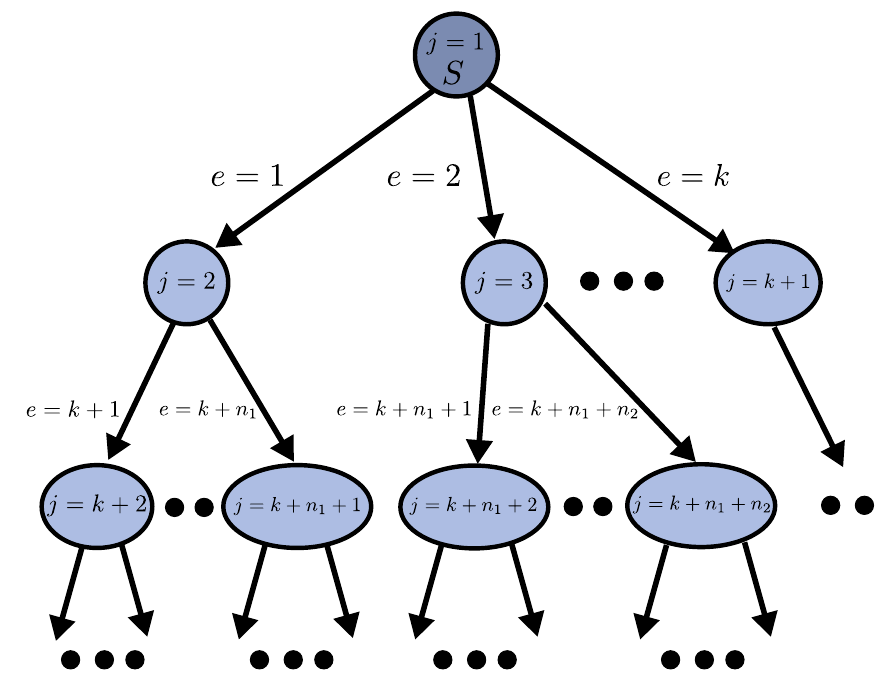}
    \end{center}
    \caption{Labeling of nodes (blue circles) and edges (black arrows) in a tree network used in Sec.~\ref{sec:lossy-tree}. The slack node is taken as the root of the tree and labeled as $j=1$ as indicated by the letter $S$ and the darker blue colouring.
    }
    \label{fig:tree}
\end{figure}

We first introduce the basic notation, see Fig.~\ref{fig:tree}. The slack node is chosen to be the root of the tree and labeled as $j=1$. The remaining nodes are labeled according to the distance to the root: first nearest neighbors, then next-to-nearest neighbors, and so on. Every edge $e=1,\ldots,M=N-1$ points to the node $e+1$. For each node and edge, we must keep track of how it is connected to the root of the tree. We thus introduce the matrix $\boldsymbol T \in \mathbb{R}^{M \times M}$ by
\begin{align*}
    T_{e,j} = 
    \begin{cases}
    +1 & \parbox[t]{.3\textwidth}{if edge $e$ is on the path from node $j+1$ to the root} \\
    0  & \mbox{otherwise}.
    \end{cases}
\end{align*}
Note that the edges are labeled in such a way that $T_{e,k}$ also indicates whether edge $e$ is on the path from edge $k$ to the root. Furthermore, we introduce the vectorial notation 
\begin{align*}
    \vec F &= (F_1,\ldots, F_M)^\top, \\
    \vec L &= (L_1,\ldots, L_M)^\top, \\
    \vec x &= (F_1,\ldots, F_M,L_1,\ldots, L_M)^\top.
    \end{align*}
    The dynamic condition  (\ref{eq:dyncon}) then reads
\begin{equation}
  \vec P  = \boldsymbol \II  \vec x,
  \label{eq:dyncon2}
\end{equation}
where the matrix $\boldsymbol \II \in \mathbb{R}^{(N-1) \times 2M}$ is obtained by concatenating the signed and unsigned incidence matrix $(\boldsymbol I \, | \, \boldsymbol E)$ and removing the first line corresponding to the slack node. In particular, the matrix elements are given by 
\begin{equation}
   \II_{j-1,e} = \left\{
   \begin{array}{r l}
      +1 & \; \parbox[t]{.3\textwidth}{if $e \le M$ and $j$ is the tail of edge $e$ or if $e > M$ and $j$ is the tail or head of edge $e-M$}  \\
      -1 & \parbox[t]{.3\textwidth}{if $e \le M$ and $j$ is the head of edge $e$}  \\
      0     & \parbox[t]{.3\textwidth}{otherwise}.
  \end{array} \right.
\end{equation}

\begin{figure}[tb]
	\centering
	\begin{center}
	   \includegraphics[width=\columnwidth]{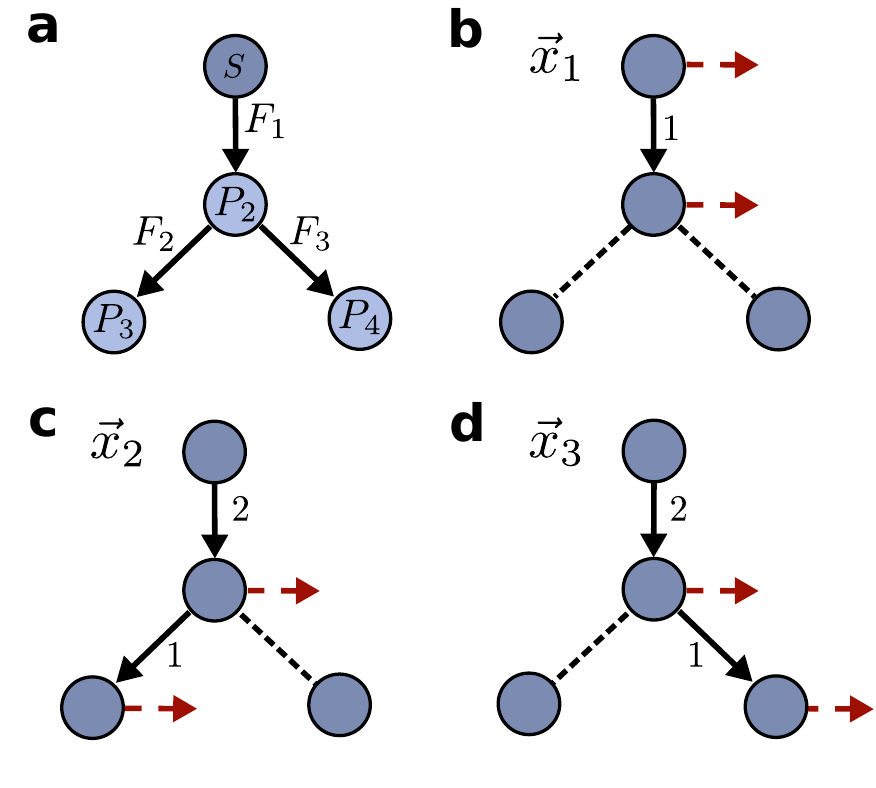}
	\end{center}
	\caption{(a) Simple tree network with $N=4$ nodes $M=3$ edges. Arrows indicate the orientation of edges which in turn determines the direction of flows. (b-d) Illustration of the basis vectors of the kernel of the matrix $\boldsymbol \II$.  The vectors $\vec x_{e}, e=1,2, 3$ include losses at exactly one edge $e$, indicated by the dotted red arrows at the terminal nodes, and the flows needed to compensate these losses. 
	}
	\label{fig:treegrid-example}
\end{figure}

First, we need a specific solution $\vec x^{(s)}$ of the dynamic condition (\ref{eq:dyncon2}). For the sake of simplicity, we choose a solution with no losses, that is
\begin{equation}
    \vec x^{(s)} = (\vec F_1^{(s)}, \ldots, \vec F_M^{(s)}, 0, \ldots 0)^\top,
    \label{eq:specific-noloss}
\end{equation} 
where 
\begin{equation}
    F_e^{(s)} = -  \sum_{j=2}^N T_{e,j-1} P_j \, .
\end{equation}
Then we have to construct the general solution to the dynamic conditions, i.e., we need a basis for the $N$-dimensional kernel of the matrix $\boldsymbol \II$. The basis vectors are constructed such that they have losses only at one particular line, which yields
\begin{align*}
     \label{eq:tree-basis-1ton}
      \vec x^{(e)} &= 
      \begin{bmatrix}
         \vec F^{(e)} \\ \vec L^{(e)} 
      \end{bmatrix},   \qquad \forall e\in\{1,\ldots,M\}   \\
     F^{(e)}_k &= 2 T_{e,k} + \delta_{e,k}  \\
     L^{(e)}_k &= \delta_{e,k} 
\end{align*} 
with the Kronecker symbol $\delta_{e,k}$. This set of basis vectors is illustrated in Fig.~\ref{fig:treegrid-example} for an elementary example. We note that these basis vectors are linearly independent as required, but not orthogonal. All solution candidates of the dynamic and the flow-loss conditions can be written as
\begin{equation}
    \vec x =  \vec x^{(s)} + \sum_{e=1}^M \alpha_e \vec x^{(e)},
\end{equation}
In terms of the flows and losses this yields
\begin{align}
    F_e &= F_e^{(s)} +
    2 \sum_{k=e+1}^{N} T_{e,k} \alpha_k + \alpha_e, \nonumber \\
    L_e &= \alpha_e.
            \label{eq:ansatzFL-tree}
\end{align}      
To simplify the notation, we introduce the abbreviation
\begin{align}
    \FF_e = -  \sum_{j=2}^N T_{e,j-1} P_j +  2 \sum_{k=e+1}^{N} T_{e,k} \alpha_k,
\end{align}
which is the flow on the line $e$ minus the losses,
\begin{align*}
    \FF_e = F_e - L_e = F_e - \alpha_e.
\end{align*}

Now we can calculate the parameters $\alpha_e$ by substituting ansatz (\ref{eq:ansatzFL-tree}) into the flow-loss condition (\ref{eq:flowloss}):
\begin{equation}
   \left( \frac{\FF_e + \alpha_e}{b_e} \right)^2 
   + \left(  \frac{\alpha_e}{g_e} -1 \right)^2 = 1.
   \label{eq:alpha_pm}
\end{equation}
To solve these quadratic equations we now have to proceed iteratively from $e=N-1$ to $e=1$ as the quantity $\FF_e$ depends on the losses $\alpha_k$ on the lines $k=e+1,\ldots,N-1$. In each step, we have to check whether the solutions are real, positive and respect the line limits (\ref{eq:linelimits}). Fortunately, these conditions can be simplified to a single inequality condition as stated in the following lemma.

\begin{lemma}
\label{lem:rlalpha}
Eq.~\eqref{eq:alpha_pm} has two real positive solutions $\alpha_e^{\pm}$ which both satisfy the line limits (\ref{eq:linelimits}) if and only if
\begin{equation}
   b_e^2 \ge \FF_e^2 + 2 g_e \FF_e.
   \label{eq:alpha-limits}
\end{equation}
The two solutions coalesce in the case of equality.
\end{lemma}

We emphasize that condition (\ref{eq:alpha-limits}) has to be satisfied for all edges $e\in\{1,\ldots,M\}$, which again has to be verified iteratively. A proof of this result is given in Appendix~\ref{sec:proof_lemma_3}. 

Finally, we can summarize our findings as follows.

\begin{lemma}
All potential solutions of the dynamic conditions and the load-flow condition for a tree network can be written as  
\begin{align*}
    F_e &= \underbrace{- \sum_{j=2}^N T_{e,j-1} P_j +  2 \sum_{k=e+1}^{N} T_{e,k} \alpha_k}_{=: \FF_e} + \alpha_e \\
    L_e &= \alpha_e,
\end{align*}
where the parameters $\alpha_e$, $e\in\{M,M-1,\ldots,1\}$
are determined iteratively as  
\begin{equation} 
   \alpha_e^{\pm} = \frac{g_e b_e}{(g_e^2 + b_e^2)}
   \left[ b_e - \frac{g_e}{b_e} \FF_e
   - \sigma_e
   \sqrt{b_e^2 - \FF_e^2 - 2 g_e \FF_e } \right],
   \label{eq:alpha-tree}
\end{equation}
where the sign $\sigma_e \in \{-1,+1\}$ indicates the solution branch. Hence, each potential solution is uniquely characterized by the sign vector $\vec \sigma = (\sigma_1,\ldots,\sigma_M)^\top \in \{-1,+1\}^M$.  
\end{lemma}

\subsection{Example}
\begin{figure*}[tb!]
    \begin{center}
        \includegraphics[width=\textwidth]{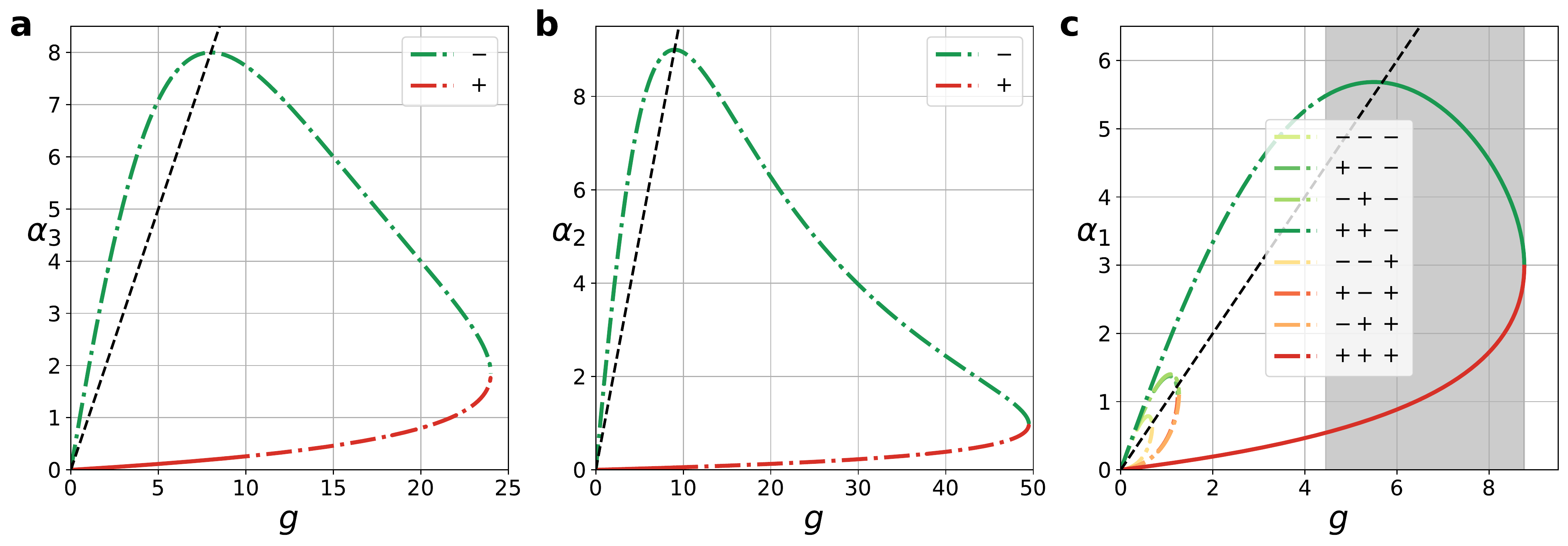}
    \end{center}
    \caption{Multiple solutions in a tree network with Ohmic losses. The possible values of the line losses $\alpha_e$ are shown as a function of the conductance $g$ for the simple four node tree network shown in Fig.~\ref{fig:treegrid-example} and parameters $b = 10$, $P_2 = -1$, $P_3 = -1$ and $P_4 = -2$ for varying $g$ as calculated according to Eq.~\eqref{eq:alpha-exc}.  Solid, coloured lines indicate dynamically stable solutions and dotted, coloured lines indicate unstable ones. The black dotted line indicates points with $\alpha_e =g$, thus determining which branch to choose when calculating angular differences according to Eq.~\eqref{eq:Delta-FL-loss}. (a,b) Branching of $\alpha_{2}$ and $\alpha_3$ into two different solutions according to the different signs of the square root in the expression (Eq.~\eqref{eq:alpha-exc}).
    (c) The solutions found for $\alpha_2$ and $\alpha_3$ can be used to subsequently calculate the solutions for $\alpha_1$. The solutions depend on the signs $\sigma_e$ for all lines $e=1,2,3$ such that we find $2^3$ solution branches in total. The signs indicated in the legend are ordered as $(\sigma_3,\sigma_2,\sigma_1)$. In the region shaded in grey, there are two coexisting \emph{stable} solutions.
    }
    \label{fig:alpha_vs_g_tree}
\end{figure*}

As an example we consider a grid with $N=4$ nodes and $M=3$ edges as depicted in Fig.~\ref{fig:treegrid-example} (a). The node-edge incidence matrix $I$ and its \emph{modulus} $E$ read
\begin{equation*}
    \boldsymbol I =\begin{bmatrix}
    +1 & 0 &  0\\ 
    -1 & +1 & +1 \\
    0 &  -1 & 0 \\
    0 & 0 & -1
    \end{bmatrix}
    \Rightarrow 
    \boldsymbol E = 
    \begin{bmatrix}
    +1 & 0 &  0\\ 
    +1 & +1 & +1 \\
    0 &  +1 & 0 \\
    0 & 0 & +1
    \end{bmatrix},
\end{equation*}
and the tree matrix is given by
\begin{equation*}
   \boldsymbol T = 
    \begin{bmatrix}
    +1 & +1 & +1\\ 
    0 & +1 & 0 \\
    0 &  0 & +1
    \end{bmatrix} \, .
\end{equation*}
The dynamic condition (\ref{eq:dyncon}) thus reads
\begin{equation*}
    \begin{cases}
    P_2 = - F_1 + F_2 + F_3 + L_1 + L_2 +  L_3\\
    P_3 = - F_2 - L_2 \\
    P_4 = - F_3 - L_3
    \end{cases}.
\end{equation*}
A particular solution of these equations is given by
\begin{equation*}
   \vec x^{(s)} = 
   \begin{bmatrix}
    \vec F^{(s)} \\ \vec L^{(s)}
    \end{bmatrix} = 
   (-P_2-P_3-P_4,-P_3,-P_4,0,0,0)^\top .
\end{equation*}
and the kernel is spanned by the basis vectors 
\begin{align*}
    \vec{x}_1 &= (1,0,0,1,0,0)^\top, \\
    \vec{x}_2 &= (2,1,0,0,1,0)^\top, \\
    \vec{x}_3 &= (2,0,1,0,0,1)^\top,
\end{align*}
which are illustrated in  Fig.~\ref{fig:treegrid-example} (b-d). Hence, the general solution can be written as 
\begin{equation*}
    \vec x = 
    \begin{bmatrix}
    F_1\\F_2\\F_3\\L_1\\L_2\\L_3
    \end{bmatrix}=\begin{bmatrix}
    F_1^{(S)}+2\alpha_3+2\alpha_2+\alpha_1\\
    F_2^{(S)}+\alpha_2\\
    F_3^{(S)}+\alpha_3\\
    \alpha_1\\\alpha_2\\\alpha_3
    \end{bmatrix}.
\end{equation*}
The coefficients $\alpha_{e}$, $ e \in\{1,2,3\}$, are directly calculated in the order $e=3,2,1$ via Eq.~\eqref{eq:alpha-tree}
with $\FF_3 = F_3^{(s)}$, $\FF_2 = F_2^{(s)}$ and $\FF_1 = F_1^{(s)}+2\alpha_2^{\pm}+2\alpha_3^\pm$. We recall that in contrast to the cyclic case we do not have to consider the geometric condition. The values of $\alpha_e^\pm$ and hence also the flows and losses depend only on the signs $(\sigma_1,\sigma_2,\sigma_3)$ -- and of course on the system parameters.

To explore the emergence of multistability in networks with Ohmic losses, we plot the different solution branches as a function of the conductances $g$ in Fig.~\ref{fig:alpha_vs_g_tree}. For the sake of simplicity we assume that all lines have the same parameters, and keep both $b$ and the power injections fixed. 
In the limit $g \rightarrow 0$, we trivially have $\alpha_e = 0$ for all edges such that the functions $\alpha_e^+$ and  $\alpha_e^-$ coalesce. However, this does not imply that equilibria themselves coalesce, cf.~Eq.~\eqref{eq:Delta_pm}. For small values of $g$, the line losses $\alpha_e$ then increase approximately linearly and we find $2^3$ different solutions in total, corresponding to the different choices of the signs $(\sigma_1,\sigma_2,\sigma_3)$. For each edge, the $+$ branch corresponds to a solution with low losses $L_e < g_e$ and the $-$  branch to a solution with high losses $L_e > g_e$. Nonlinear effects become important for higher values of $g$: The losses in the $+$ branches increase super-linearly, while the $-$ branches show a non-monotonic behaviour. For even higher values of $g$ solutions vanish pairwise. The solution branches $\vec \sigma = (+,+,+)$ with the lowest overall losses and the branch $\vec \sigma = (+,+,-)$  vanishes last.

\begin{figure}[tb!]
    \centering
    \includegraphics[width=\columnwidth]{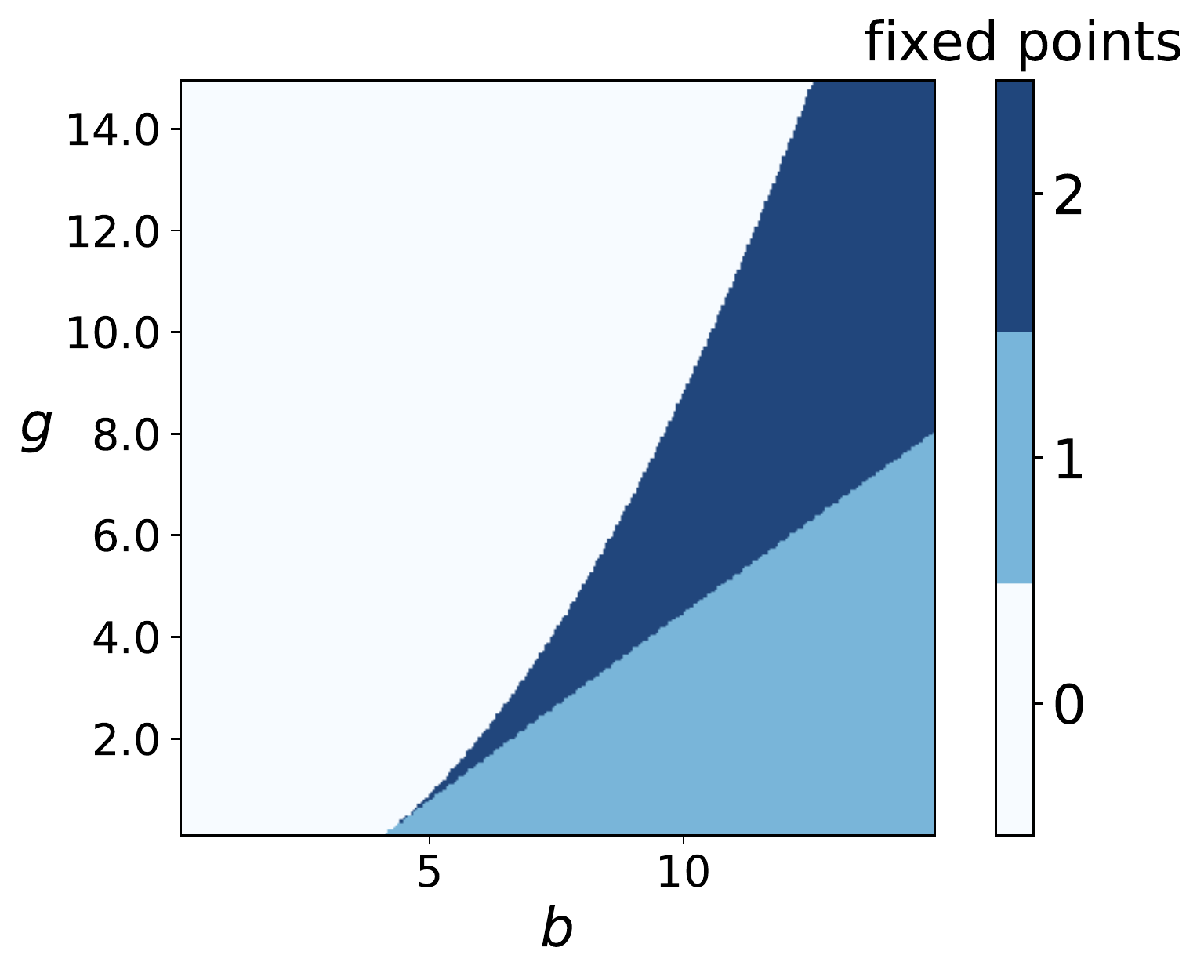}
    \caption{Number of stable fixed points (colour code) of the lossy real power flow equation~\ref{eq:realpower-intro} for the four node tree network shown in Fig.~\ref{fig:treegrid-example}(a) with power injections $P_2=-1$, $P_3=-1$ and $P_4=-2$ for varying line susceptance $b$ (abscissa) and conductance $g$ (ordinate). Whereas a minimum line capacity is required to result in any stable fixed points in the same way as for the lossless power flow, two effects that do not exist in the lossless case may be observed: Increasing conductances $g$ and thus losses requires for higher line capacities $b$ as expected. In addition to that, an additional stable fixed point arises for higher losses thus presenting a different mechanism for multistability.
    }
    \label{fig:stabilitymap_tree}
\end{figure}

We further evaluate the dynamical stability for each solution branch by numerically testing the eigenvalues of the matrix $\boldsymbol \Lambda$ defined in \eqref{eq:def-Laplace}. The weights used in this Laplacian  matrix can be rewritten directly in terms of the flows and losses. If nodes $j$ and $k$ are connected via edge $e$, we obtain
\begin{align*}
   w_{jk} &= \frac{b_e}{g_e} ( g_e - L_e )
            \pm \frac{g_e}{b_e} F_e,
\end{align*}
where the minus sign is chosen if $j$ is the tail and $k$ the head of edge $e$ and the plus sign is chosen if $k$ is the tail and $j$ the head of edge $e$.

The results for the stability of the different solution branches are indicated by displaying the lines as either dashed (unstable) or solid (stable) in Fig.~\ref{fig:alpha_vs_g_tree}. We find that only the $(+,+,+)$-branch is stable for low losses. This is expected because in the lossless case there can be at most one stable solution\cite{manik2017}. The $(+,+,+)$-branch continuously merges into this stable solution in the limit $g \rightarrow 0 $. More interestingly, also the $(+,+,-)$-branch becomes stable for large values of $g$. Hence, losses can stabilize fixed points.

A comprehensive analysis of the existence of solutions for the given sample network in terms of the grid parameters $b$ and $g$ is given in Fig. \ref{fig:stabilitymap_tree}. Remarkably, the presence of Ohmic losses has two antithetic effects on the solvability of the real power load-flow equations: On the one hand, losses can prohibit the existence of solutions. Real power flows are generally higher in lossy networks as losses have to be balanced by additional flows. Hence, the minimum line strength $b$ required for the existence of a solution increases with $g$. On the other hand, losses facilitate multistability. While the lossless equation can have at most one stable fixed point for tree networks, two stable fixed points can exist if losses are added.

For example, for three consumer nodes with power injections $P_2=-1,P_3=-1$ and $P_3=-2$ and uniform line parameters of $b=10$ and $g=8$, we find a dynamically stable solution branch with $\vec{\sigma}=(+,+,-)$ with flows $\vec{F}\approx(9.01, 1.04,2.2)^\top$ and losses $\vec{L}\approx(4.54,0.04,0.2)^\top$ and another one with $\vec{\sigma}=(+,+,+)$ with flows $\vec{F}\approx(6.2 ,1.04,2.2)^\top$ and losses $\vec{L}\approx(1.73,0.04,0.2)^\top$. We recall that node 1 serves as a slack node. Hence, the power injection $P_1$ (or the natural frequency $\omega_1$ in the oscillator context) is different for the two stable steady states.

\section{Cyclic network}
\label{sec:cycle}
\subsection{Fundamentals}
\label{sec:lossy-cycle}

\begin{figure}[tb!]
    \begin{center}
        \includegraphics[width=\columnwidth]{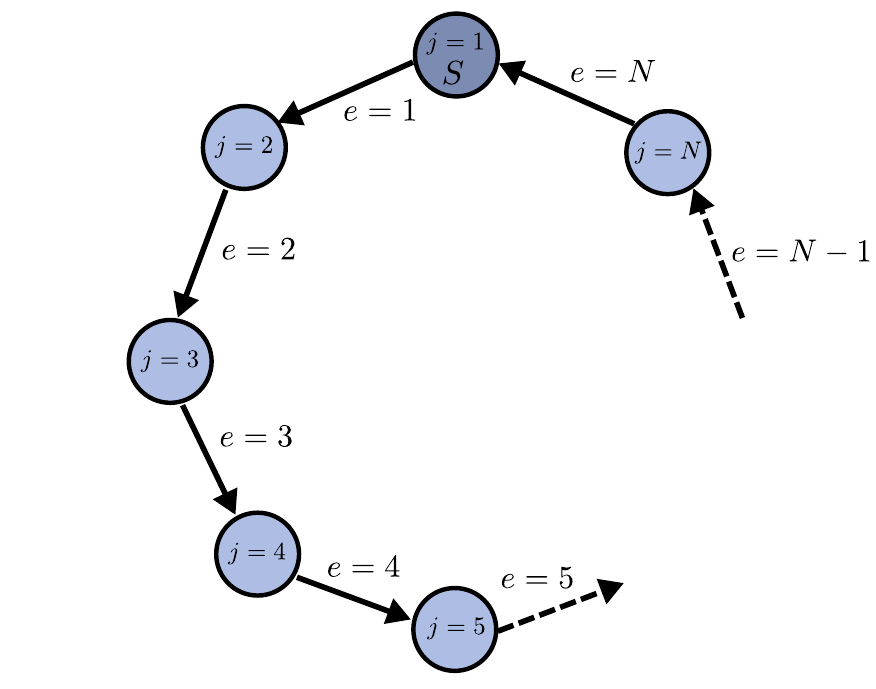}%
    \end{center}
    \caption{Labeling of nodes (blue circles) and edges (black arrows) in a cyclic network used in Sec.~\ref{sec:lossy-cycle}. The slack node is located at $j=1$ and indicated here by the letter $S$ and a colouring in darker blue.
    }
    \label{fig:cycle_label}
\end{figure}

We now consider a single closed cycle as depicted in Fig.~\ref{fig:cycle_label}. We label all nodes by $j\in\{1,\ldots,N \}$ around the cycle in the mathematically positive direction starting at the slack node $j=1$. Similarly, we label all lines $e\in\{1,\ldots,N  \}$ where line $e$ corresponds to $(e,e+1)$ and line $e=N$ corresponds to $(N,1)$. 

We now construct the solutions of the dynamic condition (\ref{eq:dyncon2}). As before, we choose a specific solution with no losses (cf.~Eq.~\eqref{eq:specific-noloss}), where the flows satisfy
\begin{equation*}
    P_j = \sum_{e=1}^N I_{j,e} F_e^{(s)}, \quad \forall j\in\{2,\ldots,N\}.
\end{equation*}
A solutions always exists as the linear set of equations has rank $N-1$. A proper initial guess can be obtained, for example, by solving the DC approximation \cite{wood2013}.

To construct the general solution, we further need a basis for the $(N+1)$-dimensional kernel of the matrix $\boldsymbol \II$. As before, we use a set of basis vectors that have losses only at one particular line, 
\begin{equation}
     \label{eq:cycle-basis-1ton}
      \vec x^{(e)} = (\underbrace{2,\ldots,2}_{e-1 \, {\rm times}},
          1, \underbrace{0,\ldots,0}_{N-e \, {\rm times}},\underbrace{0, \ldots,0}_{e-1 \, {\rm times}}, 1, \underbrace{0, \ldots,0}_{N-e \, {\rm times}})^\top.                          
\end{equation}
In contrast to the tree network we need an additional basis vector describing a cycle flow
\begin{equation}
   \label{eq:cycle-basis-np1}
     \vec x^{(N+1)} = (\underbrace{1,1,\ldots,1}_{N \, {\rm times}},
                                 0 \ldots,0)^\top.
\end{equation}

\begin{figure}[tb!]
    \begin{center}
        \includegraphics[width=\columnwidth]{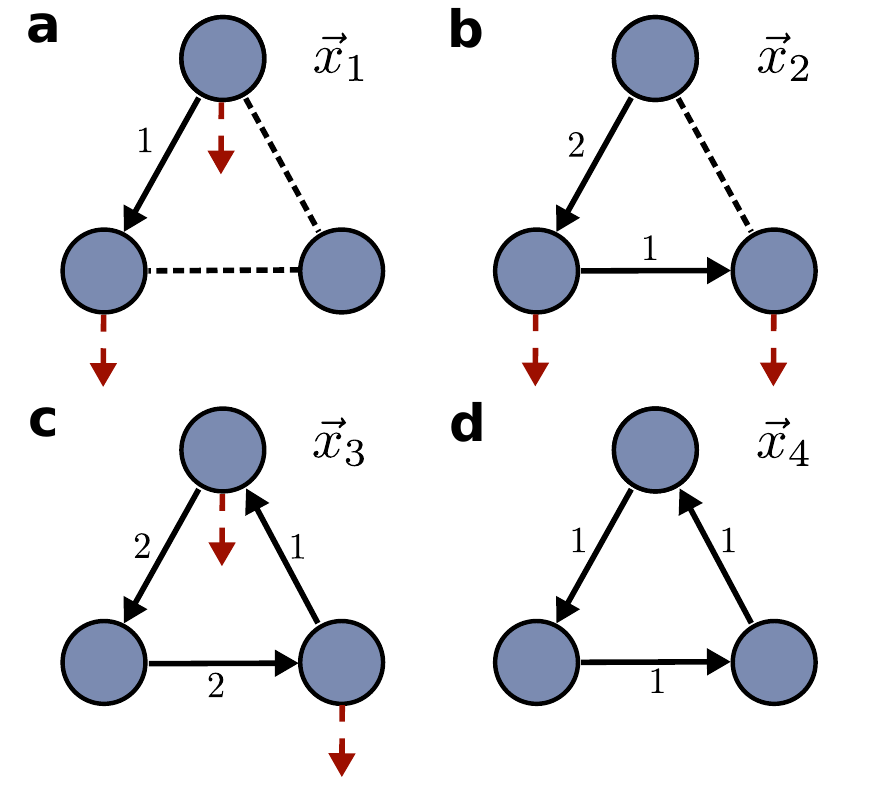}%
    \end{center}
    \caption{Illustration of the basis vectors of the kernel of the matrix $\boldsymbol \II$ for a small cyclic network with $N=3$ nodes. (a-c) The vectors $\vec x_{e}, e=1,  \ldots, N$, include losses at exactly one edge $e$, indicated by the dotted red arrows at the terminal nodes, and the flows needed to compensate this loss. (d) The basis vector $\vec x_{N+1}$ represents a lossless cycle flow.
    }
    \label{fig:basis_vectors}
\end{figure}

This set of basis vectors is illustrated in Fig.~\ref{fig:basis_vectors}. All solution candidates of the dynamic and the flow-loss conditions can thus be written as
\begin{equation}
    \vec x =  \vec x^{(s)} + f \vec x^{(N+1)} 
         + \sum_{e=1}^N \alpha_e \vec x^{(e)},
\end{equation}
where $f \in \mathbb{R}$ is a parameter giving the cycle flow strength. In terms of the flows and losses this yields
\begin{align}
    F_e &= \underbrace{F_e^{(s)} + f +
       2 \sum_{n=e+1}^{N} \alpha_n}_{=: \FF_e}
       + \alpha_e, \nonumber \\
    L_e &= \alpha_e.
        \label{eq:ansatzFL}
\end{align}      

\begin{figure}[tb]
	\centering
	\begin{center}
	   \includegraphics[width=\columnwidth]{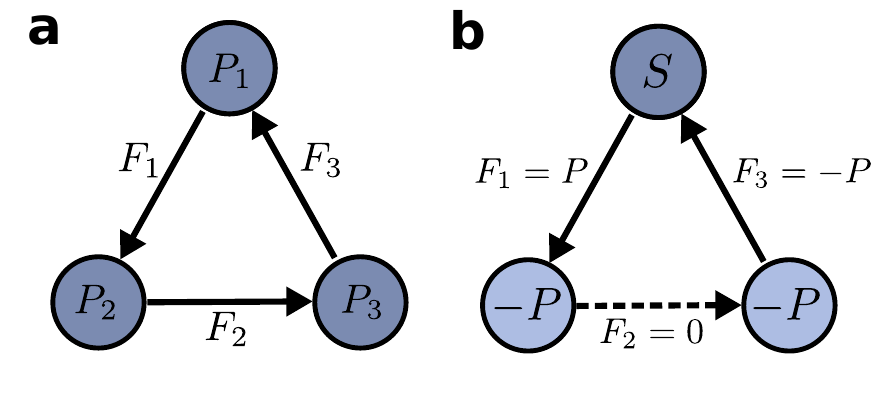}
	\end{center}
	\caption{Simple cycle network with three nodes (blue circles) and three edges (black arrows). (a) Arrows indicate the orientation of edges which in turn determines the direction of flows. We consider a network with power injections at the nodes $P_1$, $P_2$ and $P_3$ and power flows on the edges denoted $F_1$, $F_2$ and $F_3$. (b) Example studied in section \ref{sec:cycle-ex}. The node $j=1$ is chosen as a slack node and the (indicated by symbol $S$) and the two other nodes are assumed to be consumer nodes with $P_{2,3}=-P$.
	Arrows again represent the edge orientations and the values give the specific solution
	$F_1^{(s)}=P$, $F_2^{(s)}=0$ and $F_3^{(s)}=-P$.
	}
	\label{fig:threenodenetwork}
\end{figure}

As before, we can now calculate the parameters $\alpha_e$ iteratively from $e=N$ to $e=1$ using Eq.~\eqref{eq:alpha-tree}
\begin{equation}
   \alpha_e^{\pm} = \frac{g_e b_e}{(g_e^2 + b_e^2)}
   \left[ b_e - \frac{g_e}{b_e} \FF_e
   - \sigma_e
   \sqrt{b_e^2 - \FF_e^2 - 2 g_e \FF_e } \right].
   \label{eq:alpha-exc}
\end{equation}
However, we now have to take into account that the quantities $\FF_e$ also depend on the parameter $f$  -- the cycle flow strength. Hence, each potential solutions is now characterized by the continuous parameter $f$ in addition to the signs  $\sigma_1,\ldots,\sigma_N \in \{-1,+1\}$. Whether a solution exists and respects the line limits can be determined from Lemma \ref{lem:rlalpha}, in particular from condition (\ref{eq:alpha-limits}). We stress that this condition must be satisfied for all edges $e\in\{1,\ldots,N\}$ simultaneously, keeping in mind that the quantities $\FF_e$ depend on the values $\FF_{e+1},\ldots,\FF_N$ and  the cycle flow strength $f$. Hence, condition (\ref{eq:alpha-limits}) must be checked iteratively for all $e=N,N-1,\ldots,1$ in dependence of the value of $f$.

\begin{figure*}[tb]
    \begin{center}
        \includegraphics[width=\textwidth]{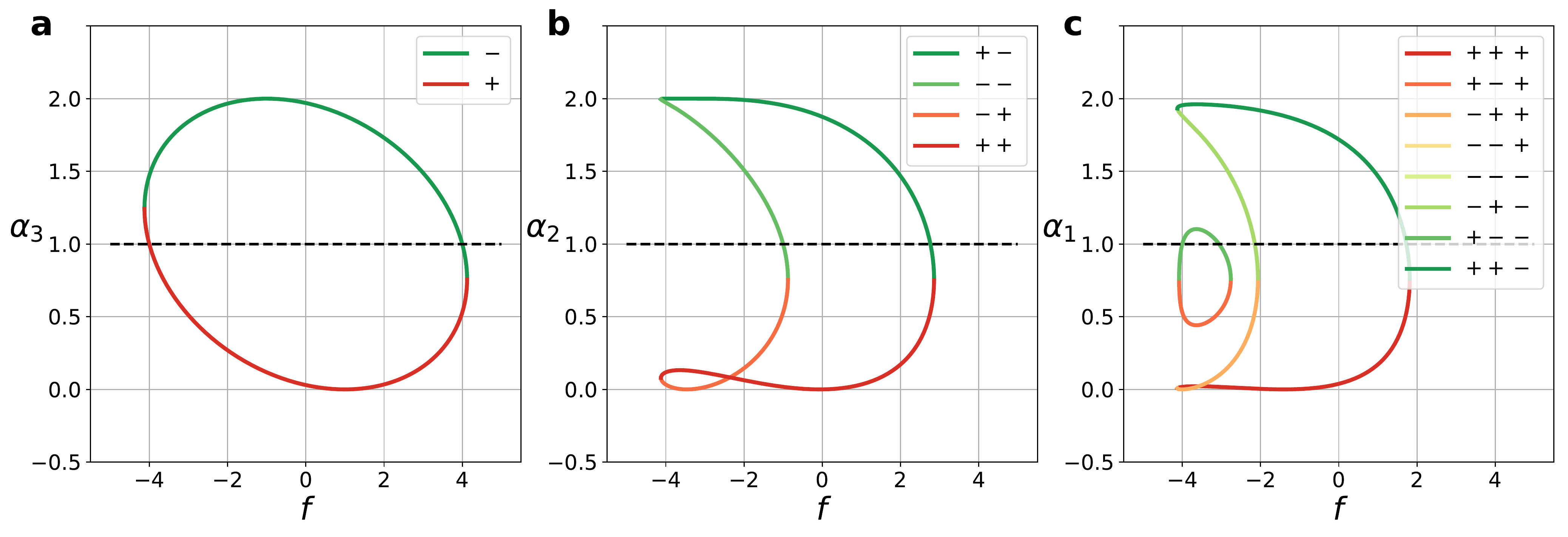}
    \end{center}
    \caption{The possible values of line losses $\alpha_{1,2,3}$ as a function of the cycle flow strength $f$ for the simple three node cycle network shown in Fig.~\ref{fig:threenodenetwork} and parameters $P=-1$, $g=1$ and $b=4$. Black dotted line indicates values where $\alpha_e=g$, thus determining the sign of angular differences according to Eq.~\eqref{eq:Delta-FL-loss}. (a) Branching of $\alpha_3$ into two different solutions referred to as $\alpha_3^+$ (dark red, bottom) and $\alpha_3^-$ (dark green, top) for the different signs of the square root as predicted by Eq.~\eqref{eq:alpha-exc}. (b-c) The solutions found for $\alpha_3$ can be used to subsequently calculate the solutions for $\alpha_2$ and then $\alpha_1$. The signs indicated here in the legend are ordered as $(\sigma_3)$, $(\sigma_3,\sigma_2)$ and $(\sigma_3,\sigma_2,\sigma_1)$ for panels (a), (b) and (c), respectively. 
    }
    \label{fig:alphas}
\end{figure*}

In a cyclic network we further have to satisfy the geometric condition (\ref{eq:geo}), which fixes the remaining continuous degree of freedom $f$. For a single cycle, the winding number is given by 
\begin{equation*}
    \varpi^{\vec \sigma} = \frac{1}{2\pi} \sum_{e=1}^M \Delta_e^{\sigma_e},
\end{equation*} 
The phase differences $\Delta_e^{\sigma_e}$ and hence the winding number are determined by the line flows and losses via Eq.~\eqref{eq:Delta-FL-loss} and depend on the respective solution branch indicated by the signs $\vec \sigma$. Recall that the geometric condition states that the winding number $\varpi$ can be an arbitrary integer. Hence there can be multiple solutions for $f$ for a given set of signs $\sigma_1,\ldots,\sigma_N$ if the cycle is large enough. This route to multistability was analyzed in detail for lossless networks in \cite{manik2017}.

\subsection{Example}
\label{sec:cycle-ex}

We analyze here a three-node cycle shown in Fig.~\ref{fig:threenodenetwork} where node $1$ is the slack node. The node-edge incidence matrix $\boldsymbol I$ and its \emph{modulus} $\boldsymbol E$ will then be
\begin{equation*}
    \boldsymbol I=\begin{bmatrix}
    +1&0&-1\\ 
    -1&+1&0 \\
    0&-1&+1
    \end{bmatrix}
    \Rightarrow 
    \boldsymbol E=\begin{bmatrix}
    +1&0&+1\\ +1&+1&0\\
    0&+1&+1
    \end{bmatrix} \, .
\end{equation*}
The dynamic condition (\ref{eq:dyncon}) thus reads
\begin{equation*}
    \begin{cases}
    P_2=F_2-F_1+L_1+L_2\\
    P_3=F_3-F_2+L_2+L_3.
    \end{cases}
\end{equation*}
A particular solution of these equations is given by
\begin{equation*}
   \vec x^{(s)} = 
   \begin{bmatrix}
    \vec F^{(s)} \\ \vec L^{(s)}
    \end{bmatrix} = 
   (-P_2,0,+P_3,0,0,0)^\top .
\end{equation*}
and the kernel is spanned by the basis vectors 
\begin{align*}
    \vec{x}_1 &= (1,0,0,1,0,0)^\top, \\
    \vec{x}_2 &= (2,1,0,0,1,0)^\top, \\
    \vec{x}_3 &= (2,2,1,0,0,1)^\top, \\
    \vec{x}_4 &= (1,1,1,0,0,0)^\top,
\end{align*}
which are illustrated in Fig.~\ref{fig:basis_vectors}. Hence, the general solution can be written as 
\begin{equation}
    \vec x = 
    \begin{bmatrix}
    F_1\\F_2\\F_3\\L_1\\L_2\\L_3
    \end{bmatrix}=\begin{bmatrix}
    F_1^{(s)}+f+2\alpha_3+2\alpha_2+\alpha_1\\
    F_2^{(s)}+f+2\alpha_3+\alpha_2\\
    F_3^{(s)}+f+\alpha_3\\
    \alpha_1\\\alpha_2\\\alpha_3
    \end{bmatrix}.
\end{equation}
The coefficients $\alpha_{1,2,3}$ are calculated as a function of $f$ iteratively starting from $N=3$ via Eq.~\eqref{eq:alpha-exc}
with $\FF_3 = F_3^{(s)}+f$, $\FF_2 = F_2^{(s)}+f + 2 \alpha^\pm_3$ and $\FF_1 = F_1^{(s)}+f+2\alpha^\pm_2+2\alpha^\pm_3$. 
The results are shown in Fig.~\ref{fig:alphas} (a-c) for all different possible realizations of the sign vector $(\sigma_1,\sigma_2,\sigma_3)$: for $\alpha_3$ we have 2 choices, then for $\alpha_2$ we have $2^2=4$ choices (two choices for each of $\alpha_2$ and $\alpha_3$) and finally we have $2^3=8$ choices for $\alpha_1$. For the sake of simplicity, we have chosen $P_2=P_3=1$ in this example. Notably, all branches of the solutions must form closed curves when plotted via the parameter $f$. This is due to the fact that a real solution of the Eq.~\eqref{eq:alpha-exc} can only vanish when the discriminant goes to zero, i.e.,~when it collides with another branch of the solution.

\begin{figure}
    \centering
    \includegraphics[width=\columnwidth]{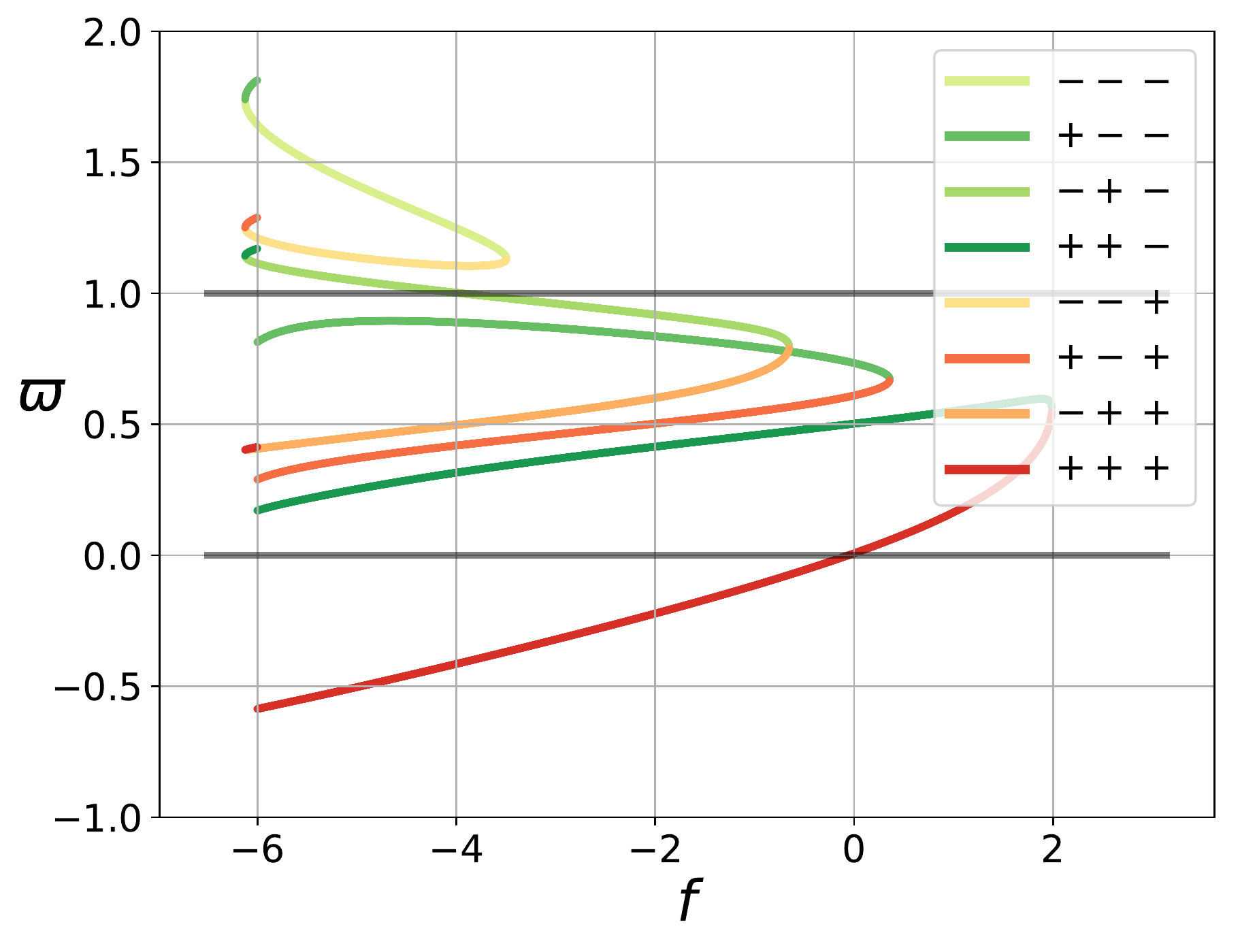}
    \caption{The winding number $\varpi$ as a function of the cycle flow strength $f$ for different solution branches in the three node network depicted in Fig.~\ref{fig:threenodenetwork}. Solutions require that $\varpi \in \mathbb{Z}$, cf.~Eq.~\eqref{eq:geo}. Colour code as in Fig.~\ref{fig:alphas},c for all panels. 
    }
    \label{fig:omega}
\end{figure}

The remaining parameter $f$ is determined by the geometric condition (\ref{eq:geo}). To evaluate this condition and to finally determine all steady states we plot the winding number 
\begin{equation*}
    \varpi^{\vec \sigma}(f) = \frac{1}{2\pi} \sum_{e=1}^M \Delta_e
\end{equation*}
as a function of $f$ in Fig.~\ref{fig:omega}. The phase differences are given by (cf.~Eq.~\eqref{eq:Delta-FL-loss})
\begin{align*}
     \Delta^{\sigma_e}_{e} = \left\{ \begin{array}{l l l}
     \arcsin(F_{e}/b_{e})  & \mbox{ if } &
     L_e \le g_e \\
    \pi - \arcsin(F_{e}/b_{e}) & \mbox{ if } & 
    L_e > g_e.
    \end{array} \right.
\end{align*}
They depend on the solution branch, i.e., on the values of the $\sigma_e$ and so does the winding number. For the given cyclic network we find $2^3$ solution branches, which have to be considered when evaluating the geometric condition, see Fig.~\ref{fig:omega}. Inspecting the winding number $\varpi^{\vec \sigma}(f)$ for each branch, we find 2 steady states of which one is stable and one is unstable. Again, the stable fixed point is given by the $(+,+,+)$-branch which has the lowest Ohmic losses.

However, we can find two dynamically stable branches for higher losses as in the case for the tree network. For example, fixing line susceptances and conductances $b=g=3$ and power injections $P_2=P_3=-1$, we find again two dynamically stable branches corresponding to low losses $\vec{\sigma}=(+,+,+)$ and high losses $\vec{\sigma}=(+,+,-)$.

\section{Summary and Discussion}
In this article, we studied solutions to the real power load-flow equations in AC transmission grids of general topology with a special focus on the impact of Ohmic losses. Extending our previous work \cite{manik2017}, we constructed an analytical method for computing all load flow solutions, both stable and unstable ones. We demonstrated how to explicitly compute all steady states in two elementary test topologies: a $4$-node tree and a $3$-node ring. 

We find that analogous to the lossless case, different solutions exist corresponding to different winding numbers \eqref{eq:geom_condition} along each basis cycle, as well as a choice between two solution branches in each edge. The two branches correspond to a state with low losses and phase differences on the respective edge ($+$ branch) and high losses and phase difference ($-$ branch). 

We show that Ohmic losses have two conflicting effects on the existence and number of steady states: On the one hand, high losses must be compensated by higher flows. Hence, solutions may vanish due to Ohmic losses unless the line capacities are also increased. On the other hand, Ohmic losses can stabilize certain solution branches and thus foster multistability. In particular, we demonstrate that two grid topologies that have been proven to exhibit \emph{no} multistability in the lossless case --  trees and $3$-node rings -- \emph{are} multistable in the lossy case for certain parameter values. 

\begin{acknowledgments}

We thank Tom Brown and Johannes Schiffer for valuable discussions. We gratefully acknowledge support from the German Federal Ministry of Education and Research (grant no. 03EK3055B) and the Helmholtz Association (via the joint initiative ``Energy System 2050 -- A Contribution of the Research Field Energy'' and the grant no.~VH-NG-1025). D.M. acknowledges funding from the Max Planck Society.
\end{acknowledgments}

\appendix

\section{Proof of lemma~\ref{lem:stability_condition}}
\label{sec:proof_lemma_1}

\begin{proof}
The result can be proven by making use of Gershgorin's circle theorem~\cite{gershgorin1931uber}. Recall that the equilibrium is linearly stable if all the eigenvalues $\mu_j$ of the Laplacian $\boldsymbol \Lambda$ have a positive real part,
\begin{align*}
  \operatorname{Re}(\mu_j)>0,\forall j\in\{1,...,N-1\},
\end{align*}
except for the eigenvalue $\mu_N=0$ corresponding to a global phase shift. According to Gershgorin's theorem, each eigenvalue $\mu_j$ is located in a disk in the complex plane with radius $R_j=\sum \nolimits_{\ell\neq j} \abs{\Lambda_{j,\ell}}$ centred at $\Lambda_{j,j}$. If the condition $w_{jk}>0$ is satisfied, we have that $\abs{\Lambda_{j,\ell}}=-\Lambda_{j,\ell}$. Therefore, applying Gershgorin's theorem results in the following inequality
\begin{align*}
    \abs{\mu_j-\Lambda_{j,j}}&\leq \sum \nolimits_{\ell\neq j}\abs{\Lambda_{j,\ell}}= \Lambda_{j,j}.
\end{align*}
This inequality thus predicts that all eigenvalues $\mu_j$ have real part greater than or equal to zero $\operatorname{Re}(\mu_j)\geq 0$. 
Now it remains to show that the eigenvalue $\mu_N$ to the eigenvector $(1,1...,1)^\top$ is the only zero eigenvalue. Assume that $\vec{v}\in\mathbb{R}^N$ is an eigenvector with eigenvalue $\mu = 0$. Assume that this vector has its minimum entry at position $i$, such that $v_i=\operatorname{min}(v_j),~j\in\{1,...N\}$ and hence $v_i-v_j\leq 0,~\forall j$. Then we arrive at 
\begin{align*}
    0 &= (\boldsymbol \Lambda\vec{v})_i= \sum_{j\neq i} \Lambda_{ij}(v_i-v_j).
\end{align*}
Since the off-diagonal elements $\Lambda_{ij}$ are all negative by the assumption of the lemma, it follows that the entries of the vector at neighbouring nodes equal its minimum value $v_i=v_j$. We can now apply the same reasoning for next-nearest neighbours and proceed in the same way through the whole network to show that 
\begin{align*}
    v_i=v_j,~\forall j\in\{1,...N\},
\end{align*}
which proofs that $\vec{v}=(1,...,1)^\top$ is the only eigenvector with vanishing eigenvalue $\mu=0$.
\end{proof}

\section{Proof of lemma~\ref{lem:rlalpha}}
\label{sec:proof_lemma_3}

\begin{figure}[tb]
    \begin{center}
        \includegraphics[width=\columnwidth]{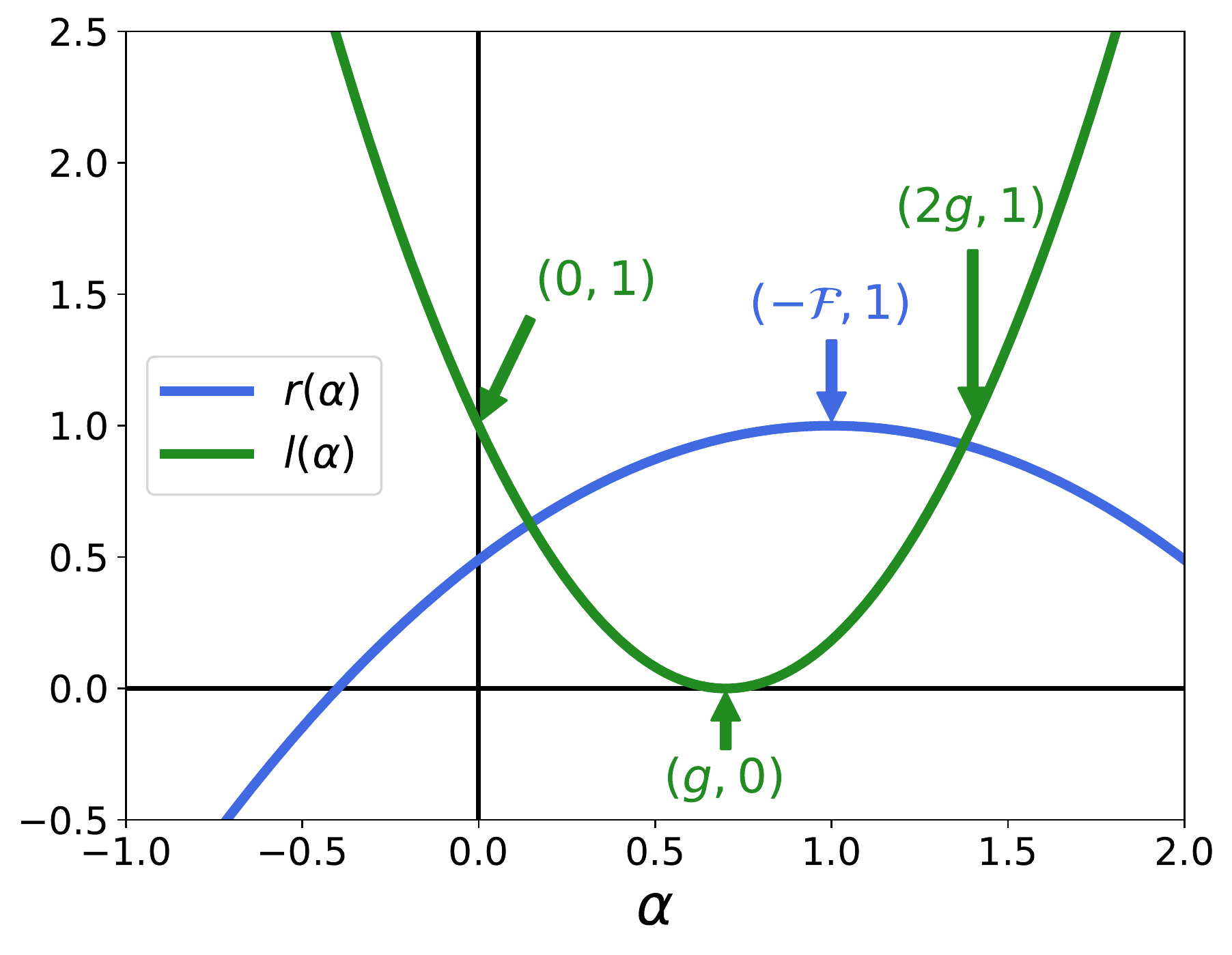}%
    \end{center}
    \caption{Illustration of the two function $l(\alpha_e)$ and $r(\alpha_e)$ used in the proof of lemma \ref{lem:rlalpha} for arbitrary parameter values $g=0.7$, $b=1.4$ and $\FF=1$.}
    \label{fig:rl-proof}
\end{figure}

\begin{proof}
We first note that if condition (\ref{eq:alpha-limits}) is satisfied, the discriminant in Eq.~\eqref{eq:alpha-tree} is non-negative, such that all solutions are real. The two solutions coalesce if the discriminant vanishes, i.e., if $b_e^2 = \FF_e^2 + 2 g_e \FF_e$. Conversely, if the condition (\ref{eq:alpha-limits}) is not satisfied, the discriminant in Eq.~\eqref{eq:alpha-tree} is negative, such that no real solution exists.

It remains to be shown that if a solution exists, then it is positive and respects the line limits. To this end, we rewrite the flow-loss condition~(\ref{eq:flowloss}) as
\begin{equation}
  \underbrace{ \left( \frac{\alpha_e}{g_e} -1  \right)^2 }_{=: l(\alpha_e)} =
  \underbrace{1 - \left( \frac{\alpha_e + \FF_e}{b_e}  \right)^2}_{=: r(\alpha_e)} %
\end{equation}
The two parabola $l(\alpha_e)$ and $r(\alpha_e)$ are illustrated in Fig.~\ref{fig:rl-proof}. The left-hand side $l(\alpha_e)$ is non-negative everywhere with
\begin{align*}
    &l(\alpha_e) \in [0,1] & \; \mbox{if}  \; \alpha_e \in [0,2g_e] \\
    &l(\alpha_e) > 1       & \; \mbox{if}  \; \alpha_e \notin [0,2g_e].
\end{align*} 
The right-hand side is smaller or equal to one with
\begin{align*}
    r(\alpha_e) \in [0,1] &\qquad \mbox{if}  \; 
        \alpha_e \in [-b_e-\FF_e,+b_e-\FF_e] \\
    r(\alpha_e) < 0        &\qquad \mbox{if}  \; 
        \alpha_e \notin [-b_e-\FF_e,+b_e-\FF_e].
\end{align*} 
Hence, we find the necessary condition for the crossing of the two parabola as
\begin{align*}
    & l(\alpha_e) = r(\alpha_e) \in [0,1], \\
    & L_e = \alpha_e \in [0,2g_e], \\
    & F_e = \FF_e + \alpha_e \in [-b_e,+b_e]. 
\end{align*}
That is, if a solution $\alpha_e$ exists, it is guaranteed to be positive and satisfy the line limits.
\end{proof}

\section*{References}
\bibliography{aipsamp.bib}

\end{document}